\documentclass[conference]{IEEEtran}
\IEEEoverridecommandlockouts
\usepackage{cite}
\usepackage{amsmath,amssymb,amsfonts}
\usepackage{makecell}
\usepackage{graphicx}
\usepackage{algorithm}
\usepackage{algpseudocode}
\usepackage{textcomp}
\usepackage{xcolor}
\usepackage{subfigure}
\usepackage{color, soul}
\usepackage{tabularx}
\usepackage[draft, commentmarkup=todo,
  todonotes={textsize=tiny, textwidth=0.83in}]{changes}

\def\BibTeX{{\rm B\kern-.05em{\sc i\kern-.025em b}\kern-.08em
    T\kern-.1667em\lower.7ex\hbox{E}\kern-.125emX}}

\usepackage[colorlinks=true,citecolor=blue,linkcolor=magenta]{hyperref}
\usepackage[]{changes}

\begin{document}

\title{Hybrid Quantum-Classical Multilevel Approach for Maximum Cuts on Graphs}

\author{\IEEEauthorblockN{Anthony Angone}
\IEEEauthorblockA{\textit{Quantum Science and Engineering} \\
\textit{University of Delaware}\\
Newark, DE \\
aangone@udel.edu}

\and

\IEEEauthorblockN{Xiaoyuan Liu}
\IEEEauthorblockA{
\textit{Fujitsu Research of America, Inc.}\\
Sunnyvale, CA \\
xliu@fujitsu.com}

\and

\IEEEauthorblockN{Ruslan Shaydulin}
\IEEEauthorblockA{
\textit{Global Technology Applied Research} \\
\textit{JPMorgan Chase}\\
New York, NY \\
ruslan.shaydulin@jpmchase.com}

\and

\IEEEauthorblockN{Ilya Safro}
\IEEEauthorblockA{\textit{Department of Computer and Information Sciences} \\
\textit{University of Delaware}\\
Newark, DE \\
isafro@udel.edu}
}

\maketitle

\begin{abstract}
Combinatorial optimization is one of the fields where near term quantum devices are being utilized with hybrid quantum-classical algorithms to demonstrate potentially practical applications of quantum computing. One of the most well studied problems in combinatorial optimization is the Max-Cut problem. The problem is also highly relevant to quantum and other types of ``post Moore'' architectures due to its similarity with the Ising model and other reasons. In this paper, we introduce  a scalable hybrid multilevel approach to solve large instances of Max-Cut using both classical only solvers and quantum approximate optimization algorithm (QAOA).
We compare the results of our solver to existing state of the art large-scale Max-Cut solvers. We demonstrate excellent performance of both classical and hybrid quantum-classical approaches and show that using QAOA within our framework is comparable to classical approaches.\\
Reproducibility: Our solver is publicly available at \url{https://github.com/angone/MLMax-cut}.
\end{abstract}

\section{Introduction}

Recently, a plethora of post-Moore hardware architectures have been developed, demonstrating promising potential in tackling combinatorial optimization problems. These include quantum annealers\cite{quantumannealing}, gate-based quantum devices which operate using fundamental quantum gates and circuits, digital and analog annealers\cite{digitalannealing}, often considered as intermediate architectures between classical and quantum computers, and  Coherent Ising Machines\cite{ising}. Some of these machines are generally called Ising processing units (IPU) due to its architecture designed to solve Ising model optimization problems. Despite the potential of these architectures, many suffer from the limitation of a small number of variables and connectivity, creating a significant barrier in addressing larger, complex real-world problems. To address these issues, hybrid approaches that blend classical and novel hardware have emerged, where the primary routine runs on a classical machine and calls the specialized hardware to solve small sub-problems, effectively combining the best of both worlds \cite{shaydulin2019hybrid}. While the experimental focus of this paper is \emph{hybrid quantum-classical and classical-classical approaches} for the maximum cut (Max-Cut) on graphs, nothing  prevents from applying a similar approach on other novel architectures.

Recent Noisy Intermediate Scale Quantum (NISQ) devices is one of these novel architectures. They suffer from having few qubits and high error rates~\cite{preskill2018quantum}. Regardless there is a tremendous interest in exploring NISQ devices for practical applications \cite{Herman2023,baiardi2022quantum}. This has encouraged the development of various hybrid algorithms that combine the use of both classical and quantum computers. Notable hybrid algorithms are variational algorithms such as the Quantum Approximate Optimization Algorithm (QAOA)~\cite{farhi2014quantum} and the Variational Quantum Eigensolver (VQE)~\cite{peruzzo2014variational}. 
These types of algorithms construct a shallow parameterized quantum circuit, which a classical optimizer then attempts to optimize over the parameters to find the optimal solution. 

Due to the engineering and physical limitations, the number of qubits available to use is too small to actually solve problems in practical applications. This situation inspires the development of using classical decomposition based techniques to decompose the original problem into subproblems that will be solved on a quantum computer, then the classical computer is used to piece these subproblems together into a solution for the original problem~\cite{shaydulin2019hybrid,ushijima2021multilevel}. If these subproblems are chosen in a way that they are all independent from each other, they can be trivially solved in parallel on quantum devices or IPUs before being pieced back together. 

The focus of this paper is the multilevel approach, one of the most significant classes of methods for large scale computationally hard problems. These algorithms look at the problem through a sequence of increasingly coarser representations of the original problem. They have proven to be efficient in many different domains such as graph partitioning \cite{karypis1995analysis} and linear arrangement \cite{SAFRO200624}.\\

\noindent {\bf Our contribution:} 

In this paper, we present a novel approach to solving the Max-Cut problem using a multilevel method. For the coarsening within our multilevel solver, we introduce a distance measure designed specifically for the Max-Cut problem that takes advantage of a $d$-dimensional embedding of the nodes. We also introduce a novel randomized, multistart refinement scheme. We evaluate our solver on a diverse set of graphs including real world problems such as social networks. We demonstrate our solver's performance both in the quantum-classical regime using QAOA as our subproblem solver and in the fully classical regime using a MQLib heuristic as the subproblem solver. Our experimental results show that our solver is competitive with state-of-the-art global solvers when provided with equivalent computational resources but can use small IPUs or universal quantum devices in parallel. We also demonstrate that our solver is scalable, being competitive both for large graphs and dense graphs.

\section{Background and Notation}

\subsection{Maximum cut on graphs}
The Max-Cut is an NP-hard combinatorial optimization problem defined on a graph. We consider simple undirected weighted graphs defined as $G = (V, E, w)$ where $V$ is a set of nodes, $E$ is a set of edges, and edge weight function 
$w:E\rightarrow \mathbb{R}_{\geq 0}$. 
We define $n=|V|$ and $m=|E|$. A cut is a partition of $V$ into two disjoint parts, $V_1$ and $V_2$. The objective to be maximized is the weighted sum of edges $ij\in E$ such that $i \in V_1$ and $j \in V_2$ or vice versa. We can formulate Max-Cut as a quadratic unconstrained binary optimization problem (QUBO) that is equivalent to the models solved by IPUs and universal quantum devices using QAOA (see next section). Binary variables $x_i$ are interpreted as follows:
\begin{equation}
    x_i =
    \begin{cases}
        0, & \text{if} ~i \in V_1 \\
        1, & \text{if} ~i \in V_2.
    \end{cases}
\end{equation}
The Max-Cut optimization problem is then defined as:
\begin{equation}\label{eq:maxcut}
    \max_{x\in \{0,1\}^n} \sum_{ij\in E} w_{ij}(x_i + x_j - 2x_ix_j).
\end{equation}

\subsection{Quantum Approximate Optimization Algorithm}

The Quantum Approximate Optimization Algorithm (QAOA), is a hybrid quantum optimization algorithm proposed in~\cite{farhi2014quantum}. This algorithm uses a parameterized quantum circuit of $p$ layers of alternating unitary operators and a classical optimization solver. The classical optimization solver  optimizes over the parameters of the circuit and tries to maximize the expectation.

Suppose we are trying to optimize some objective function $f(x)$, $x \in \{0,1\}^n$. When using QAOA to solve this, we would begin by building the parameterized quantum circuit. This quantum circuit has two parameters: $\beta$ and $\gamma$. It is constructed by applying alternating unitary operators: the phase operator $U(\gamma) = e^{-i\gamma\mathcal{H}}$ and the mixing operator $U(\beta) = e^{-i\beta B}$. $\mathcal{H}$ is the problem Hamiltonian while $B$ is the mixing Hamiltonian. Given the number of layers $p$ and parameters $\beta$ and $\gamma$, we can construct our quantum state:
\begin{equation}
    \vert \gamma,  \beta \rangle = U(\beta_p)U(\gamma_p)\dots U(\beta_1)U(\gamma_1)\vert + \rangle ^{\oplus n}
\end{equation}
We then measure the circuit and compute the objective function:
\begin{equation}
    \langle \mathcal{H} \rangle = \langle\gamma,\beta \vert \mathcal{H} \vert \gamma,  \beta\rangle
\end{equation}
The classical optimizer then attempts to find values of $\gamma$ and $\beta$ such that the objective is maximized. The state $\vert\gamma_{opt},\beta_{opt}\rangle$ corresponds to the optimal solution. There exist several versions of QAOA. In this work we implemented the classical QAOA of depth $p=3$.

\subsection{Multilevel Algorithms}

Multilevel algorithms are a powerful class of computational methods used for handling and processing large-scale instances in an efficient and manageable way. In the context of graphs, these algorithms  break down the large-scale graph problem into a series of increasingly simpler subproblems at multiple scales of coarseness, which improves scalability and for graph problems helps to take advantage of the graph structure \cite{brandt2003multigrid}. The multilevel algorithm  involves three main phases: coarsening,  coarsest solution, and uncoarsening. In the coarsening phase, the graph is successively reduced in size to create a series of smaller, simplified graphs. Once the coarsest (i.e., smallest and fast to solve) graph is reached, the coarsest solution phase begins, where the graph problem is solved either exactly or approximately but with a high quality. Following this, the solution is then progressively uncoarsened, where the solution is interpolated back to the original graph and refined at each level of coarseness. The entire multilevel approach is inspired by the multigrid methods.

The coarsening stage involves starting off with the original graph $G_0$ and constructing a sequence of $L$ increasingly coarser graphs $G_1, G_2, ... G_L$. Each next coarser graph $G_k$ has fewer nodes than the previous coarse graph $G_{k-1}$. To denote the sets of nodes and edges of graph at level $k$ we will use notation $G_k = (V_k, E_k)$ for $0\leq k \leq L$. To describe the coarsening from level $f$ to level $c = f+1$, we will use subscript letters $c$ and $f$ to define coarse and fine variables, respectively.

The coarsening process is performed by grouping nodes into pairs that form next-coarser nodes.  As a result of this, the edges are also merged into next-coarser level edges and the edge weights are accumulated. 

The uncoarsening stage involves using the solution from a coarser problem as an initialization to begin solving a finer problem and refinement. At the initialization step, each of the nodes is uncoarsened into the two nodes that were merged into the coarse node during the coarsening stage. The values of the variables are then carried from the coarse variables to the fine variables. After the initialization is finished, the refinement is started in which a local small sub-problem solver is used to improve upon the solution that was carried over in iterative manner. This process repeats at each level of coarseness until we return to the original graph that we started with before any coarsening.

\subsection{Multistart Methods}
When solving combinatorial problems, it is very common to get trapped in a local optima that prevents achieving the global optima. One technique that has been used to improve solutions beyond a local optima is multistart methods \cite{marti2013multi,shaydulin2019multistart}. Many optimization techniques consist of starting at some initial solution and then improving upon that solution. Multistart methods improve upon this by trying multiple initial solutions, improving all of them, and then accepting the best of the resulting solution.

\section{Related Work}
\subsection{Coarsening}
When designing a multilevel algorithm for some problem, an important factor is the coarsening scheme. When working with the Max-Cut problem, the coarsening needs to be different than the one usually designed for the graph cut or edge length minimization problems \cite{karypis1995analysis,SAFRO200624,safro2006multilevel,walshaw2004multilevel} due to the  maximization objective. Typical coarsening methods for cut minimization use either  regular algebraic multigrid (AMG) methods or restricted AMG (in which only pairs of nodes are coarsened that are faster and more popular) where a coarse subset of the graph nodes is chosen to form a coarse node based on the connectivity strength between nodes. The intuitive idea is to coarsen the heavy edges and leave sparse cuts to solve at the coarse levels. When working with the Max-Cut problem, coarsening nodes that are strongly connected does not help since we would lose heavy edges that span between the two parts. Instead, we designed a coarsening technique that coarsens together nodes that are loosely connected. This results in losing fewer edges when coarsening which means that the progressive change in the Max-Cut value will not be  significant throughout the hierarchy.
\subsection{QAOA Acceleration}
Our algorithm  relies on using QAOA to solve subproblems. Thus its refinement performance is dependent on that of QAOA. There have been many attempts to improve the performance of QAOA including parameter transferring \cite{transferability23}, warm-starts \cite{egger2021warm}, phase operator sparsification \cite{liu2022quantum} and taking advantage of symmetry \cite{shaydulin2021classical}.
\subsection{Solving Large Max-Cut Instances}
With our algorithm, we attempt to solve large Max-Cut problems. There have been many attempts at this previously, including tabu search \cite{kochenberger2013solving}, semi-definite programming relaxations \cite{grippo2012speedp}, and randomized algorithms \cite{festa2002randomized}. Tabu search is a type of algorithm that is similar to local search algorithms. Differences include that it may accept a worse solution to escape a local optima and it will mark some solutions to not be visited again. This method has been used to solve many well known graphs of sizes $|V|=800...10000$ to the best known solutions \cite{Maxcut2013}.  Semi-definite programming relaxations have also been used as a way to approximate problems that have a linear or convex quadratic objective. Randomized algorithms are also a technique commonly used in many combinatorial optimization problems as a simple and efficient way to improve upon other algorithms. When using local search algorithms, adding random perturbations enables the algorithm to escape a local optima. One widely known algorithm that utilizes both of these techniques is the Goemans-Williamson algorithm which is able to approximate the solution to a ratio of 0.878\cite{GW95}. \\
\subsection{MQLib}
Throughout this paper, we frequently reference MQLib\cite{DunningEtAl2018} which is a project that implemented and evaluated 37 heuristics for Max-Cut and QUBO problems. One of the heuristics that was implemented is a rank-two relaxation heuristic presented in \cite{burer2002}. We use this implementation both as a subproblem solver within our algorithm and as a global solver to compare our algorithm against. This specific heuristic was chosen due to it performing the best in 5 of the 7 metrics considered in MQLib's systematic evaluation of the 37 heuristics and performed the best on our benchmark that contains graphs larger than those in MQLib.

\section{Hybrid Quantum-Classical Multilevel Solver for Max Cut}

We introduce a novel multilevel hybrid quantum-classical algorithm designed for the Max-Cut problem. The  algorithm is tailored not only to leverage the capabilities of quantum architectures, but it is also adaptable to general Ising Processing Units (IPUs), devices designed for solving  combinatorial optimization problems, in which they serve as efficient sub-problem solvers \cite{coffrin2019evaluating}. To the best of our knowledge, this is the first multilevel Max-Cut algorithm. Inspired by relaxation-based algebraic multigrid coarsening techniques \cite{ron2011relaxation,livne2012lean}, our proposed algorithm introduces two principal innovative components. The first is the novel relaxation-based Max-Cut distance measure to inform and optimize the coarsening phase designed in the spirit of algebraic distance on graphs \cite{chen2011algebraic}. The second is a fast, multistart refinement.

\subsection{Coarsening}

As has been shown in many multigrid and multilevel algorithms
, a distance measure between variables is one of the most critical components of the coarsening \cite{ron2011relaxation}. Our distance measure between nodes 
uses a $d$-dimensional  embedding space to pair together nodes to be coarsened. This process begins by creating a $d$-dimensional sphere that will have every node embedded into it. Initially, every node starts off at a random position in the space
\begin{equation}
    \forall i\in V ~~~ p_i^{(0)} \leftarrow \text{rand}[-1,1]^d.
\end{equation}
Each node in the graph is iteratively visited and its position is updated in a way that maximizes the total distance between the node and every other node in its neighborhood.
\begin{equation}\label{eq:maxdist}
\forall i\in V ~~~ p_i^{(l+1)}\leftarrow \max \sum_{j\in N(i)}w_{ij}||p_i^{(l)} - p_j^{(l)}||_2,
\end{equation}
where $l$ is the number of the iteration, and $N(i)$ is the set of neighbors for $i\in V$. This process in Eq. (\ref{eq:maxdist}) is repeated until convergence. Once the required number of iterations are completed, the embedding process is finished and the nodes can start to be paired together. A similar (but  faster to compute) distance measure in which the embedding space was a $d$-dimensional hypercube led to similar results.

\begin{algorithm}
\caption{Coarsening}\label{alg:coarsening}
\begin{algorithmic}[1]
\Require Fine graph $G_f = (V_f, E_f)$
\State $A_f\leftarrow$Adjacency matrix of $G_f$
\State $embedding \leftarrow embed(G_f)$
\State $used \leftarrow \emptyset$
\State $pairs \leftarrow \emptyset$
\For{$i \in V_f$}
\If{$i \not\in used$}
\State $j \leftarrow nearestNeighbor(i, embedding), \not\in used$
\State $used \leftarrow i, j$
\State $pairs \leftarrow (i, j)$
\EndIf
\EndFor
\State $P \in 0^{|V_c| \times |V_f|}$
\State $q \leftarrow 0$
\For{$(i,j) \in pairs$} 
\State $P_{q,i} \leftarrow 1$
\State $P_{q,j} \leftarrow 1$
\State $q = q + 1$
\EndFor
\State $A_c \leftarrow P^TA_fP$\\
\Return $G_c$ obtained from $A_c$
\end{algorithmic}
\end{algorithm}

When the nodes are being paired together to be coarsened, each node should be paired to a node that lies near it in the embedding. This is done by placing all the nodes and their positions in a KDTree. We then iterate through each node and perform a nearest neighbor search within the KDTree. Each node is then paired to the nearest neighbor that has not already been paired with a different node. Once this matching is constructed, the new coarse graph is built by contracting every pair in the matching. The details are in Algorithm \ref{alg:coarsening}.%\is{add this algorithm}

In line 2 of the algorithm, we embed the nodes into our embedding space according to \ref{eq:maxdist}. In lines 5-11, we iterate through each node in the graph, pairing it with the nearest node that is not yet paired. On line 12, we construct a matrix $P$ that is filled with zeros. On lines 14-18, we update the matrix $P$ so that each column contains two entries of 1 that correspond to a pair of nodes that will be merged together. In line 19, we find the adjacency matrix of our coarse graph by computing the matrix multiplication $P^TA_fP$. The function returns the coarse graph that corresponds to the coarse adjacency matrix we computed.

This coarsening process is repeated until the desired coarsest graph size is reached. Overall, it is expected that the original graph will be broken into approximately a logarithmic number of increasingly coarse graphs.

\subsection{Sparsification}
During the coarsening process, we provide an optional sparsification parameter in order to reduce the density of our coarse graphs using a novel technique based upon our embedding. This sparsification happens after the embedding but before the coarsening. Once we calculate our embedding, we compute the weighted length of each edge in that embedding. Longer edges are more likely to be within the cut, so we choose to sparsify the shortest edges in that embedding. Instead of just removing the edge, we move the weight of that edge to its longest length adjacent edge. We repeat this process until we remove the number of edges specified by a parameter inputted by the user.

\subsection{Coarsest level solution}
The size of the coarsest level is based upon the subproblem size that is acquired via user input. The graphs will get coarsened until the size of the coarsest graph is less than the subproblem size. Once the coarse graph size reaches the desired size, we solve it using MQLib with a running time of 5 seconds.

\subsection{Uncoarsening}
We begin the uncoarsening phase with an initial solution to our coarsest level. 
When we move to the next level, we inherit our solution from the previous coarsest level. This inheritance is done with a surjection $F: V_f \rightarrow V_c$ that maps from the fine nodes to the coarse nodes that they were contracted into. 
\begin{equation}
    \forall i \in V_f ~~~ x_i = x_{F(i)}
\end{equation}

At each level, we begin by computing the gain of each node, which is the change in objective if you were to move that node to the other part.

This gain can be computed as follows:
\begin{equation}
    \forall i \in V ~~~ gain(i) \leftarrow \sum_{j\in N(i)}w_{ij}(-1)^{2x_ix_j - x_i - x_j}
\end{equation}
Throughout the refinement we efficiently keep track of each node's gain by updating it based on the edges that enter/leave the cut. We solve each level by iteratively improving upon the solution until we stop all improvement. Each iteration chooses nodes to use in a subproblem, constructs a subproblem, and then solves that subproblem. This process repeats until we perform 3 iterations without improvement. The details of this are described in Algorithm \ref{alg:refinement}. In each iteration, we first choose a random subset of the nodes and order them by their gain. From that subest, we choose the $K$ nodes with the highest gain for the subproblem and then construct and solve it. We take the solution if our objective stays the same or increases. After 3 iterations with no improvement, we consider our refinement as completed and accept the current solution. We use multistart methods here to run multiple instances of this process in parallel and accept the best of those solutions.

\subsection{Subproblem construction}
Each subproblem is constructed by creating a graph with 2 super-nodes and all the nodes chosen. Each super-node is an aggregation of every node not chosen for the subproblem in each part. Edges are added between the super-nodes and regular nodes.
Once the subproblem is constructed, we will solve it using a subproblem solver. The subproblem solver used is either QAOA or a MQLib solver. The resulting solution will be no worse than our current solution. If we find a better solution, we update all the nodes to be in the correct part and update their gain accordingly.\\

\begin{algorithm}
\caption{Refinement}\label{alg:refinement}
\begin{algorithmic}
\Require A graph $G = (V, E)$
\Require Subproblem Size $K$
\Require Initial Solution $S$
\State $constructGain()$
\State $count \leftarrow 0$
\State $obj \leftarrow calculateObjective(S)$
\While{$count < 3$}
\State $subset \leftarrow$ random $max(0.2|V|,2K)$ nodes $\in V$
\State $subprob \leftarrow$ $K$ highest gain nodes $\in subset$
\State $constructSubproblem()$
\State $solveSubproblem()$
\State $S \leftarrow updateSolution()$
\State $newObj \leftarrow updateObjective()$
\State $updateGain()$
\If{$newObj > obj$}
\State $obj \leftarrow newObj$
\State $count \leftarrow 0$
\Else 
\State $count \leftarrow count + 1$
\EndIf
\EndWhile\\
\Return $S$

\end{algorithmic}
\end{algorithm}

\section{Computational Experiments}
We evaluated our solver on a set of graphs to evaluate both our performance when working with both a quantum and classical subproblem solver. \\

\subsection{Quantum Subproblem Solver}
We evaluated our solver using a quantum subproblem solver on 5 small graphs taken from the well known Gset graph set acquired from SuiteSparse Matrix Collection\cite{suitesparse}. These graphs all contain 800 nodes and 19716 edges. Our subproblem solver uses a simulated QAOA circuit with a size of 12 and a depth of 3 layers. These parameters were chosen due to the limitations of simulation. We perform the QAOA experiments on a smaller set of graphs due to the long running time of simulating QAOA. In Table \ref{table:qaoa}, we compare against the exact solutions obtained with Gurobi solver and against our own solver with a classical subproblem solver.

\begin{table}[htbp]
  \caption{QAOA Results}
  \centering
  \label{table:qaoa}
  \begin{tabular}{|l|c|c|c|c|c|}
    \hline
    G & $|V|$ & $|E|$ & QAOA ML & MQLib ML & Exact solution \\
    \hline
    G1 & 800 & 19716 & 11442 & 11424 & 11624 \\
    \hline
    G2 & 800 & 19716 & 11430 & 11413 & 11620\\
    \hline
    G3 & 800 & 19716 & 11407 & 11426 & 11622\\
    \hline
    G4 & 800 & 19716 & 11386 & 11403 & 11646\\
    \hline
    G5 & 800 & 19716 & 11429 & 11412 & 11631\\
    \hline
  \end{tabular}
\end{table}

\subsection{Classical Subproblem Solver}
We evaluated our solver with a classical subproblem solver on 30 different graphs of different types that were either constructed or taken from SuiteSparse Matrix Collection or the Network Repository\cite{nr}. The types of graphs used include social networks, random regular graphs, computational fluid dynamics problems, optimization problems, and finite graphs. The random regular dense graphs, titled r-xxxx-xxxx, are the only ones that were constructed. These experiments were performed with 40 parallel refinements in the multistart scheme and with a subproblem size of 100. In Table \ref{table:results}, we compare against using MQLib as a global solver with time constraints equivalent to the time for refinement when using our algorithm. We also provide a comparison between our solver's coarsest level objective and our solver's finest level objective.

As it can be seen from the results, the performance of our multilevel solver is at least comparable to those of the best global heuristic from MQLib and is noticeably better on some graphs. Since our refinement is decomposition based and can run in parallel, the scalability of our solver is very good. It is important to mention a remarkable quality of the coarsening that actually solves a large portion of the problem (see column ``Coarse Ratio'') as well as the sparsification that almost does not change the overall quality of our solver. 

\begin{table*}[ht]
  \centering
  \caption{Experimental Results. Columns $d_{min}$, $d_{max}$, and $d_{avg}$ correspond to the minimum, maximum and average degree in the graph, respectively. In column ``ML/MQLib Ratio'' we present the ratio of Max-Cuts obtained by our multilevel algorithms and the solution obtained from MQLib. The column ``Coarse Ratio'' shows the part of our final cut obtained by the coarsening, i.e., measured at the coarsest level. In column ``10\% Sparse Ratio'' we compare the ratio of Max-Cuts obtained by our multilevel algorithm with 10\% of edges sparsified at each level with the solution obtained from MQLib.}
  \label{table:results}
  \begin{tabular}{|l|c|c|c|c|c|c|c|c|c|}
    \hline
    $G$ & $|V|$ & $|E|$ & $d_{min}$ & $d_{max}$ & $d_{avg}$ &\makecell{ML/MQLib} & \makecell{Coarse \\Ratio} & \makecell{10\% Sparse \\ML/MQLib} \\
    \hline
    soc-brightkite & 56739 & 212945 & 1 & 1134 & 7.506 & 1.017 & 0.943 & 1.013\\
    \hline
    soc-epinions & 26588 & 100120 & 1 & 443 & 7.531 &  1.005 & 0.948 & 1.008\\
    \hline
    soc-slashdot & 70068 & 358647 & 1 & 2507 & 10.237 & 1.003 & 0.841 & 1.003\\
    \hline
    soc-buzznet & 101163 & 2763066 & 1 & 64289 & 54.626 & 1.000 & 0.863 & 1.000\\
    \hline
    r-1000-5000 & 5000 & 2500000 & 1000 & 1000 & 1000 & 1.000 & 0.981 & 1.000\\
    \hline
    r-2000-10000 & 10000 & 10000000 & 2000 & 2000 & 2000 & 1.000 & 0.990 & 1.000\\
    \hline
    r-3000-15000 & 15000 & 22500000 & 3000 & 3000 & 3000 & 1.000 & 0.990 & 1.000\\
    \hline
    r-4000-20000 & 20000 & 40000000 & 4000 & 4000 & 4000 & 1.000 & 0.993 & 1.000\\
    \hline
    r-5000-25000 & 25000 & 62500000 & 5000 & 5000 & 5000 & 1.000 & 0.992 & 1.000\\
    \hline
    3dtube & 45330 & 1629474 & 10 & 2364 & 70.894 & 1.008 & 0.815 & 0.985\\
    \hline
    copter2 & 55476 & 407714 & 4 & 45 & 13.699 & 0.990 & 0.877 & 0.990\\
    \hline
    aug2dc & 30200 & 40000 & 1 & 4 & 2.649 & 0.971 & 0.952 & 0.970\\
    \hline
    aug3d & 24300 & 34992 & 1 & 6 & 2.880 & 0.999 & 0.974 & 0.997\\
    \hline
    big\_dual & 30269 & 44929 & 2 & 3 & 2.968 & 0.985 & 0.946 & 0.984\\
    \hline
    biplane-9 &21701 & 42038 & 2 & 4 & 3.874 & 0.972 & 0.956 & 0.978\\
    \hline
    shock-9 & 36476 & 71290 & 2 & 4 & 3.908 & 0.971 & 0.955 & 0.974\\
    \hline
    rajat06 & 10922 & 28922 & 1 & 62 & 4.301 & 0.996 & 0.962 & 0.996\\
    \hline
    rajat07 & 14842 & 39342 & 1 & 72 & 4.306 & 0.989 & 0.961 & 1.001\\
    \hline
    rajat08 & 19362 & 51362 & 1 & 82 & 4.309 & 0.994 & 0.959 & 0.990\\
    \hline
    rajat09 & 24482 & 64982 & 1 & 92 & 4.312 & 0.994 & 0.966 & 0.993\\
    \hline
    rajat10 & 30202 & 80202 & 1 & 102 & 4.314 & 0.986 & 0.951 & 0.992\\
    \hline
    c-59 & 41282 & 260909 & 2 & 3090 & 11.640 & 1.034 & 0.978 & 1.026\\
    \hline
    dixmaanl & 60000 & 179999 & 4 & 5 & 4.999 & 0.990 & 0.960 & 0.990\\
    \hline
    dtoc & 24993 & 34986 & 1 & 4 & 2.799 & 0.990 & 0.949 & 0.971\\
    \hline
    ex3sta1 & 16782 & 347890 & 9 & 339 & 40.459 & 0.996 & 0.949 & 0.996\\
    \hline
    c-62 &  41731 &  300537 &  2 &  5060 &  13.403 & 1.023 & 0.971 & 1.023\\
    \hline
    c-64 & 51035 & 384438 & 2 & 9474 & 14.066 & 1.012 & 0.980 & 1.012\\
    \hline
    c-68 & 64810 & 315408 & 2 & 2774 & 8.733 & 1.074 & 0.948 & 1.074\\
    \hline
    c-71 & 76638 & 468096 & 2 & 6720 & 11.216 & 1.069 & 0.971 & 1.069\\
    \hline
    c-72 & 84064 & 395811 & 2 & 5172 & 8.417 & 1.063 & 0.950 & 1.061\\
    \hline

  \end{tabular}
\end{table*}

\subsection{Obstacles and Future Work}
There are still a few obstacles we have ran into with this solver. One obstacle we face is the density of the coarse graphs. As the size of our coarse graph decreases, our density increases significantly due to our coarsening attempting to preserve edges. Currently we attempt to alleviate this problem by utilizing sparsification but a better solution is necessary. We face another obstacle of our subproblem selection relying heavily on randomness. \\
In the future, we plan on working improving the running time of QAOA. Some ways to achieve this include taking advantage of parameter transferring \cite{transferability23} and symmetry \cite{shaydulin2021classical}. Improving the running time of QAOA will both improve the overall running time of our solver and will allow us to perform experiments on larger graphs while using QAOA.

\section{Conclusion}
With the current state of quantum computing hardware, an important problem is making practical use of these devices even with high error rates and low qubit counts. Hybrid algorithms such as QAOA have been in development to address this problem. In this paper, we presented a multilevel algorithm alongside QAOA to solve the NP-Hard Max-Cut problem. We introduced novel schemes for coarsening and refinement for the Max-Cut problem. We performed a numerical study that shows we are at least comparable (and sometimes better) against state-of-the-art solvers on a number of different graphs when ran for the same amount of time. We concluded by demonstrating that we are scalable and can solve large, practical problem instances.

\bibliographystyle{plain}
\bibliography{ref}

\section*{Acknowledgements}
This work was supported  in part with funding from the Defense Advanced Research Projects Agency (DARPA). The views, opinions and/or  findings expressed are those of the author and should not be interpreted as representing the official views or policies  of the Department of Defense or the U.S.Government.

This research was supported in part through the use of DARWIN computing system: DARWIN - A Resource for Computational and Data-intensive Research at the University of Delaware and in the Delaware Region, which is supported by NSF Grant \#1919839.

This material is based upon work supported by the University of Delaware Graduate College through the Unidel Distinguished Graduate Scholar Award. Any opinions, findings, and conclusions or recommendations expressed in this material are those of the author(s).

This paper was prepared for informational purposes with contributions from the Global Technology Applied Research center of JPMorgan Chase \& Co. This paper is not a product of the Research Department of JPMorgan Chase \& Co. or its affiliates. Neither JPMorgan Chase \& Co. nor any of its affiliates makes any explicit or implied representation or warranty and none of them accept any liability in connection with this paper, including, without limitation, with respect to the completeness, accuracy, or reliability of the information contained herein and the potential legal, compliance, tax, or accounting effects thereof. This document is not intended as investment research or investment advice, or as a recommendation, offer, or solicitation for the purchase or sale of any security, financial instrument, financial product or service, or to be used in any way for evaluating the merits of participating in any transaction.

R.S.~was supported in part by the U.S.\ Department of Energy (DOE), Office of Science, Office of Advanced Scientific Computing Research AIDE-QC and FAR-QC projects, and by the Argonne LDRD program under contract number DE-AC02-06CH11357. 

\end{document}